\documentstyle[epsf]{aipproc}

\begin{document}
\hfill{UTTG-05-01}
\title{Conference Summary\\20th Texas Symposium on Relativistic Astrophysics}

\author{Steven Weinberg}
\address{Department of Physics, University of
Texas at Austin\\weinberg@physics.utexas.edu}

\maketitle

\begin{abstract} 
This is the written version of the summary talk given at the 20th Texas Symposium on Relativistic Astrophysics in Austin, Texas, on December 15, 2000.  After a brief summary of some of the highlights at the 
conference, comments are offered on three special topics: theories 
with 
large additional spatial dimensions, the cosmological constant 
problems, and the 
analysis of fluctuations in the cosmic microwave background.
\end{abstract}

\section{OVERVIEW}

Speaking as a particle physicist, an outsider, I have to say that my 
chief 
reaction after a week of listening to talks at this meeting is one of 
envy.  
You 
astrophysicists are blessed with enlightening data in an abundance that 
particle 
physicists haven't seen since the 1970s.  And although you still face 
many 
mysteries, theory is increasingly converging with observation.

For instance, as discussed by Shri Kulkarni, it now seems clear that 
gamma ray 
bursters are at cosmological distances, producing over $10^{50}$ ergs 
in particle kinetic energies alone in a minute or so, making 
them the most spectacular objects in the sky.  Tsvi Piran described a 
fireball 
model for the gamma ray bursters, in which gamma rays are produced by 
relativistic particles accelerated by shocks within material that is 
ejected 
ultra-relativisticaly from a central source.  One can think of 
various 
mechanisms 
for the hidden central source, but even without a specific model for 
the source, 
the fireball model does a good job of accounting for what is 
observed.  

According to this fireball model, gamma rays from the bursters are 
strongly 
beamed.  Peter H\"{o}flich presented evidence that core collapse 
supernova are 
also highly aspherical.  Both conclusions may be good news for 
gravitational 
wave astronomers --- only aspherical explosions can generate 
gravitational 
waves.

Spectacular things seem to be turning up all over.  
Amy Barger told us how X-ray 
observations are revealing many active galactic nuclei in what had 
previously 
seemed like ordinary galaxies, and John Kormendy reported evidence 
that the 
events that produce galactic bulges or elliptical 
galaxies are the same as those 
that produce black holes in galactic centers.

Astrophysics is currently the beneficiary of massive surveys that are 
providing 
or will soon provide a flood of important data.  We heard from John 
Peacock 
about the 2dF Galaxy Redshift Survey, from Bruce Margon about the 
Sloan Digital 
Sky Survey, and from George Ricker about the HETE x-ray and $\gamma$-
ray 
satellite mission.  Together with cosmic microwave background 
observations, 
about which more later, there seems to be a general consistency with 
the big 
bang cosmology, with about 30\% of the critical mass furnished by 
cold dark 
matter, and about $70\%$ furnished by negative-pressure vacuum 
energy.

This is not to say that there are no puzzles.  Alan Watson reported 
on the 
long-standing puzzles of understanding how the highest energy cosmic 
rays are 
generated and how they manage to 
get to earth through the cosmic microwave background.  
There are also persistent 
problems in matching the cold dark matter model to observations of 
the mass 
distribution in galaxies.  Ben Moore cast some doubt on 
whether cold dark matter 
really leads to the missing ``cuspy cores'' of galaxy haloes, and he 
concentrated 
instead on a different problem: cold dark matter models give much 
more
matter in satellites of galaxies than is observed.  
He suggests that the missing 
satellites may really be there, and that they have not been observed 
because 
they have not formed stars.  The reionization processes discussed 
here by Paul 
Shapiro may be responsible for the failure of star formation. 

Our knowledge of the dark matter mass distribution within galaxies is 
receiving 
important contributions from observations of the lensing of quasar 
images by 
intervening galaxies, discussed by Genevieve Soucail.  There have 
been hopes of 
using surveys of gravitational lenses to distinguish among 
cosmological models, 
but I have the impression that the study of galactic 
lenses will turn out to be  
more important in learning about the lensing galaxies themselves.  
Andrew Gould 
reported that microlensing observations have ruled out the dark 
matter being 
massive compact halo objects with masses in the range $10^{-
7}M_\odot$ to $10^{-
3}M_\odot$.
 
It would of course be a great advance if cold dark matter particles 
could be 
directly detected.  We heard a lively debate about whether weakly 
interacting 
massive dark matter particles have already been detected, between 
Rita Bernabei 
(pro) and Blas Cabrera (con).  It would be foolhardy for a theorist 
to try to 
judge this issue, but at least one gathers that, if the dark matter 
is composed 
of WIMPs, then they can be detected.  

I have now completed my 10 minute general summary 
of the conference.  There were 
other excellent plenary talks, and I have not mentioned any of the 
parallel 
talks, but what can you do in 10 minutes?  In the remaining 35 
minutes, I want 
to take up some special topics, on which I will have a few comments 
of my own.

\section{LARGE EXTRA DIMENSIONS} 

It is an old idea that the four spacetime dimensions in which we live 
are 
embedded in a higher dimensional spacetime, with the extra dimensions 
rolled up 
in some sort of compact manifold with radius $R$.  This would have 
profound 
cosmological consequences: the compactification of the extra 
dimensions could be 
the most important event in the history of the universe, and 
such theories would 
contain vast numbers of new types of particle.

In the original version of 
this theory any field would have normal modes that would be observed 
in four 
dimensions as an infinite tower of `Kaluza--Klein recurrences,' 
particles 
carrying the quantum numbers of the fields, with masses given by 
multiples of 
$1/R$.  It had generally been supposed that $R$ would be of the order 
of the 
Planck length, or perhaps 10 to 100 times  larger, of the order of 
the inverse 
energy $M$  at which the strong and electroweak coupling constants 
are unified.  
Even setting this preconception aside, it had seemed that in any case 
$R$ would 
have to be smaller than $10^{-16}\;{\rm cm}\approx (100\;{\rm 
GeV})^{-1}$, in 
order that the Kaluza--Klein recurrences of the particles of the 
standard model 
would be heavy enough to have escaped detection.  

The possibilities for higher dimensional theories became much richer 
with the 
increasing attention given to the idea that the spacetime in which we 
live does 
not merely {\rm appear} four-dimensional --- our three-space may be a 
truly 
three-dimensional 
surface that is embedded in a higher dimensional space.  (This is 
the picture of 
higher dimensions that was vividly 
described in Edwin Abbott's 1884 novel {\em Flatland}, and has more 
recently 
become an 
important part of string theory, starting  with Polchinski's work on 
D-branes[1].)  This idea opens up the possibility that some fields 
may depend 
only on 
position on the four-dimensional spacetime surface, while others 
`live in the 
bulk' --- that is, they depend on position in the full 
higher-dimensional
space.  
Only the fields that live in the bulk would have Kaluza--Klein 
recurrences.

Craig Hogan here discussed the recently proposed idea that the 
compactification 
scale $R$ may actually be much larger than $10^{-16}$ cm, with no 
Kaluza--Klein 
recurrences for the particles of the standard model because the 
standard model 
fields depend only on position in the four-dimensional spacetime in 
which we 
live[2].  According to this idea, it is only the gravitational field 
that 
depends 
on position in the higher dimensional space, and so it is only the 
graviton 
that 
has has Kaluza--Klein recurrences, which at ordinary energies would 
interact 
too 
weakly to have been observed.  The long range forces produced by 
exchange of 
these massive gravitons would be small enough to have escaped 
detection in 
measurements of gravitational forces between laboratory masses as 
long as $R<1$ 
mm.  (There are stronger astrophysical and cosmological bounds on 
$R$, arising 
from limits on the production of graviton recurrences in 
supernovas[3] and in 
the early  universe[4].)

In any such theory with large compactification radius $R$ the Planck 
mass scale 
of the higher dimensional theory of gravitation would be very much 
less 
than the 
Planck mass scale in our four dimensional spacetime.  In a world with 
$4+N$ 
spacetime dimensions the gravitational constant $G_{4+N}$ (the 
reciprocal 
of the 
coefficient of the term $\int 
d^{4+N}x\,\sqrt{g}\,g_{\mu\nu}R^{\mu\nu}$ in the 
action) has 
dimensionality $[{\rm mass}]^{-2-N}$, so we would expect it to be 
given 
in terms 
of some fundamental higher dimensional Planck mass scale $M_*$ by $ 
G_{4+N}\approx M_*^{-2-N}$.  Dimensional analysis then tells us that 
the 
gravitational constant $G$ in four spacetime dimensions must be given 
by
\begin{equation}
G\approx M_*^{-2-N}R^{-N}\;.
\end{equation} 
The usual assumption in theories with extra dimensions has been that 
$R\approx 
M_*^{-
1}$, in which case $G\approx M_*^{-2}$, and $M_*$ would have to be 
about 
$10^{19}$ GeV.
But if we take $N=1$ and $R\approx $ 1 mm, then $1/R\approx 10^{-13} 
$ GeV, and 
$M_*\approx 10^8$ GeV.  With $N=2$ and $R\approx $ 1 mm, 
$M_*\approx 300$ GeV.  This is the most attractive aspect of theories 
with 
large 
extra  dimensions: they can reduce or eliminate what had seemed like 
a huge gap 
between the characteristic energy scale of electroweak symmetry 
breaking 
and the 
fundamental energy scale at which gravitation becomes a strong 
interaction.

Theories with large extra dimensions are very ingenious, and they may 
even be 
correct, but I am not enthusiastic about them, for they give up the 
one solid 
accomplishment of previous theories that attempt to go beyond the 
standard 
model:  the renormalization group equations of the original standard 
model 
showed that there is an energy, around $10^{15}$ GeV, where the three 
independent gauge coupling constants become 
nearly equal[5].  In the supersymmetric version 
of the standard model the convergence of the couplings with each 
other becomes 
more precise[6], and the energy scale $M_U$ of this unification moves 
up to 
about 
$2\times 10^{16}$ GeV [7], which is less than would be expected in 
string 
theories 
of gravitation by a factor of only about 20.  (This is also a 
plausible energy 
scale for the violation of lepton number conservation that may be 
showing up in 
the neutrino oscillation experiments discussed here by Masayuki 
Nakahata.)  The 
Kaluza--Klein tower of 
graviton recurrences does nothing to change the running of the strong 
and 
electroweak coupling 
constants, and since the higher dimensional Planck mass $M_*$ is very 
much less 
than $10^{15}$ GeV in theories with large extra dimensions (this, 
after all, is 
the point of these theories), it appears that {\em in these theories 
the 
standard model gauge couplings 
are not unified at the fundamental mass scale $M_*$.}  Of course, 
they might be 
unified at some higher 
energy, but we have no way to calculate what happens in these 
theories at any 
energy higher than $M_*$.  

In his talk here Hogan mentioned that Dienes, Dudas, and 
Gherghetta[8] have 
proposed a way out of this problem.  I looked up their papers, and 
found that 
they modify the renormalization group equations for the gauge 
couplings of the 
standard model by allowing the gauge and Higgs fields (and perhaps 
some fermion 
fields) to depend on position in the higher dimensional space, along 
with the 
gravitational field.  Of course, then they have to avoid conflict 
with 
experiment by taking $1/R$ greater than 100 GeV.  The Kaluza-Klein 
recurrences 
of the gauge bosons greatly increase the rate at which the coupling 
constants of 
the standard model run, but with little change in their unification.  
To put 
this quantitatively, Dienes {\em et al.} find the bare (Wilsonian) 
couplings 
evaluated with a cut-off $\Lambda$ are
\begin{equation}
\frac{4\pi}{g_i^2(\Lambda)}= \frac{4\pi}{g_i^2(m_Z)}-
\frac{b_i}{2\pi}\ln\frac{\Lambda}{m_Z}+\frac{\bar{b}_i}{2\pi}\ln
\Lambda R
-\frac{\bar{b}_iX_N}{2\pi N}\left[(\Lambda R)^N-1\right]\;,
\end{equation} 
where  $g_1$ and $g_2$ are defined as usual in terms of the electron 
charge $e$ 
and the electroweak mixing angle $\theta$ by $g^2_1=e^2/\sin^2\theta$ 
and 
$g^2_2=5e^2/3\cos^2\theta$; $g_3$ is the coupling constant of quantum 
chromodynamics; and $X_N$ is a number of order unity. 
The constants $(b_1,b_2,b_3)$ are the factors $(33/5, 1,-3)$ 
appearing in the 
renormalization group equation of the supersymmetric standard model 
with two 
Higgs doublets, while the constants $(\bar{b}_1,\bar{b}_2,\bar{b}_3)$ 
are the 
corresponding factors $(3/5, -3, -6)$ in the renormalization group 
equations for 
$\Lambda$ above the compactification scale $1/R$ (with a possible 
constant added 
to each of the $\bar{b}_i$, proportional to the number of chiral 
fermions that 
live in the bulk).  Dienes {\em et al.} remark that the 
standard model couplings still come close to converging to a common 
value, 
because the ratios of the differences of the $\bar{b}_i$ are not very 
different 
from the ratios of the differences of the $b_i$.  I would like to put 
this more 
quantitatively, by asking what value of $\sin^2\theta$ is needed in 
order for 
the couplings to become exactly equal at some value of $\Lambda$.
In the supersymmetric standard model, this is
\begin{equation}
\sin^2\theta=\frac{3(b_3-b_2)+5(b_2-b_1)e^2/g_3^2}{8b_3-3b_2-
5b_1}=\frac{1}{5}+\frac{7}{15}\frac{e^2}{g_3^2}=0.231\;,
\end{equation}
in excellent agreement with the measured value $0.23117\pm 0.00016$.
(Here  $e$ and $g_3$ are taken as measured at $m_Z$, in which case 
$e^2/4\pi=1/128$ and $g_3^2/4\pi=0.118$.)  If all the running of the 
couplings 
were at scales greater than $1/R$, then $\sin^2\theta$ would be given 
by
Eq.~(3), but with $b_i$ replaced with $\bar{b}_i$:
\begin{equation}
\sin^2\theta=\frac{3(\bar{b}_3-\bar{b}_2)+5(\bar{b}_2-
\bar{b}_1)e^2/g_3^2}{8\bar{b}_3-3\bar{b}_2-
5\bar{b}_1}=\frac{3}{14}+\frac{3}{7}\frac{e^2}{g_3^2}=0.243\;.
\end{equation}
This is not bad, but nevertheless outside experimental bounds.  (It 
would be 
necessary to consider higher-order contributions in the 
renormalization group 
equations and threshold effects to be sure that there is really a 
discrepancy 
here.)  In order not to 
spoil the prediction for $\sin^2\theta$, $1/R$ would have to be 
considerably 
larger than 
1 TeV, so  that much of the running of the coupling constants would 
occur at 
scales below $1/R$, where the renormalization group equations are 
those of the 
supersymmetric standard model.  

In any case, the running of the couplings is so rapid above the 
compactification 
scale $1/R$ that the couplings become equal (to the extent that they 
do become 
equal) at an energy not far above $1/R$.  The $4+N$ dimensional 
Planck scale 
$M_*$ given by Eq.~(1) is very much greater than this.  Taking $1/R$ 
greater 
than 1 
TeV, Eq.~(1) would give $M_*$ greater than $10^{13}$ GeV for $N=1$.  
Even for 
$N=7$, we would have 
$M_*$ greater than $10^{6}$ GeV.  {\em Thus theories of this sort 
save the 
 unification of couplings at the cost of reintroducing a large gap 
between 
the higher-dimensional Planck scale $M_*$ and the electroweak scale.}

\section{VACUUM ENERGY}

There are now two problems surrounding the energy of empty space[9].  
The first 
is 
the old problem, why the vacuum energy density is so much smaller 
than any one 
of a  number of individual contributions.  For instance, it is  
smaller 
than the 
energy density in quantum fluctuations of the gravitational field at 
wavelengths 
above the Planck length by a factor of about $10^{-122}$ and it is 
smaller than 
the latent heat associated with the breakdown of chiral symmetry in 
the strong 
interactions by a factor about $10^{-50}$.  All these contributions 
can be 
cancelled by just adding an appropriate cosmological constant in the 
gravitational field equations; the problem is why there should be 
such a 
fantastically well-adjusted cancellation.  The second, newer, problem 
is why the 
vacuum energy density that seems to be showing up in supernova 
studies of the 
redshift-distance relation (reviewed in a parallel session by Nick 
Suntzeff and 
Saul Perlmutter) is of the same order of magnitude (apparently larger 
by a 
factor about 2) as the matter density {\em at the present time.}  
There are five
broad classes of attempts to solve one or both of these problems:

\vspace*{10pt}

\noindent
1) {\em Cancellation Mechanisms}\\
It has occurred to many theorists that the gravitational effect of 
vacuum energy 
might be wiped out by the dynamics of a scalar field, which 
automatically 
adjusts itself to minimize the spacetime curvature.  So far, this has 
never 
worked.  Some recent attempts were described by Andre Linde in a 
parallel 
session.

\vspace*{10pt}

\noindent 
2) {\em Deep Symmetries}\\
There are several symmetries that could account for a vanishing 
vacuum energy, 
if they were not broken.  One is scale invariance; another is 
supersymmetry.  
The problem is to see how to preserve the vanishing of the vacuum 
energy despite 
the breakdown of the symmetry.  No one knows how to do this.

\vspace*{10pt}

\noindent 
3) {\em Quintessence}\\
It is increasingly popular to consider the possibility that the 
vacuum 
energy is 
not constant, but evolves with the universe[10].  For instance, a 
real scalar 
field 
$\phi$ with Lagrangian density $-\partial_\mu\phi\partial^\mu\phi/2-
V(\phi)$ if 
spatially homogeneous contributes a vacuum energy density and a 
pressure
\begin{equation} 
\rho =\frac{1}{2}\dot{\phi}^2+V(\phi)\;,~~~~~~~ p 
=\frac{1}{2}\dot{\phi}^2-
V(\phi)\;,
\end{equation} 
so the condition $\rho+3p<0$ for an accelerating expansion is 
satisfied if the 
field $\phi$ is evolving sufficiently slowly so that 
$\dot{\phi}^2<V(\phi)$.  

It must be said from the outset that, in themselves, quintessence 
theories do 
not help with the first problem mentioned above --- they do not 
explain why 
$V(\phi)$ does not contain an additive constant of the order of 
$(10^{19}{\rm 
GeV})^4$.  It is true that superstring theories naturally lead to 
``modular'' scalar fields $\phi$ for which $V(\phi)$ does vanish as 
$\phi\rightarrow \infty$, in which case the vacuum becomes 
supersymmetric.  It might be hoped that the vacuum energy is small 
now, because the scalar field is well on its way toward this limit.  
The trouble is that the vacuum now is nowhere near supersymmetric, so 
that in these theories we would expect a present vacuum energy of the 
order of the fourth power of the supersymmetry-breaking scale, or at 
least $(1\;{\rm TeV})^4$.

On the other hand, such theories may help with the 
second problem, if 
the quintessence energy is somehow related to the energy in matter 
and 
radiation,
because  the present moment  is not so many $e$-foldings of cosmic 
expansion 
(about 10, in fact) from the turning point in cosmic history when the 
radiation 
energy density (including neutrinos) fell below the matter energy 
density.  Paul 
Steinhardt here described a model in which the quintessence energy 
density was 
less than the radiation energy density by a constant factor $r$, as 
long as 
radiation dominated over matter[11].  (It is necessary that $r$ be 
considerably 
less 
than unity, in order that quintessence should not appreciably 
increase the 
expansion rate during the era of nucleosynthesis, increasing the 
present helium 
abundance above the observed value.)  Then when the radiation energy 
density 
fell below the matter energy density at a cross-over redshift 
$z_C\approx 3000$  
the quintessence energy dropped sharply by a factor of order $r^2$, 
and has 
remained roughly constant since then.  Since the cross-over between 
radiation 
and matter dominance the matter energy density has decreased by a 
factor $z_C^{-
3}$, so the ratio of the quintessence energy density and the matter 
energy 
density now should be of order $r^2\times r\times z_C^3=(z_Cr)^3$.  
For the 
quintessence and matter energies to be about equal now, $r$ must be 
equal to 
about $1/z_C\approx 3\times 10^{-4}$.  Steinhardt tells me that when 
these 
calculations are done carefully, the required ratio $r$ of 
quintessence to 
radiation energy density at early times is about $10^{-2}$, 
rather than $3\times 
10^{-4}$.  But whatever the value of $r$ that makes the quintessence 
energy 
comparable to the matter energy density now, it requires some fairly 
fine 
tuning: changing $r$ by a factor $10$ would change the ratio of the 
present 
values of the quintessence energy density and the matter energy 
density by a 
factor $10^3$. 

\vspace*{10pt}

\noindent
4) {\em Brane Solutions}

\noindent
Several authors have found solutions of brane theories of the 
Randall--Sundrum 
kind[2] in which our four-dimensional spacetime is flat, despite the 
presence of 
a large cosmological constant in the higher dimensional gravitational 
Lagrangian[12].  These solutions contained an unacceptable 
essential singularity 
off the brane, but there are models in which this can be avoided[13].  
I don't 
believe that there is anything unique in these solutions, so that 
instead of 
having to fine tune parameters in the Lagrangian one has to fine tune 
initial 
conditions.  Also, it is not clear why the effective cosmological 
constant has 
to be zero {\em now}, rather than before the spontaneous
breakdown of the chiral 
symmetry of quantum chromodynamics, when the latent heat associated 
with this 
phase transition would have given the vacuum an energy density $(1\; 
{\rm 
GeV})^4$.

\vspace*{10pt}

\noindent
5) {\em Anthropic Principle}\\
Why is the temperature on earth in the narrow range where water is 
liquid?  One 
answer is that otherwise we wouldn't be here.  This answer makes 
sense only 
because there are many planets in the universe, with a wide range of 
surface 
temperatures.  Because there are so many planets,  it is natural that 
some of 
them should have liquid water, and of course it is just these planets 
on which 
there would be anyone to wonder about the temperature.  In the same 
way, if our 
big bang is just one of  many big bangs, with a wide range of vacuum 
energies, 
then it is natural that some of these big bangs should have a vacuum 
energy in 
the narrow range where galaxies can form, and of course it is just 
these big 
bangs in which there could be astronomers and physicists wondering 
about the 
vacuum energy.  
To be specific, a constant vacuum energy if negative would have to be 
greater 
than about $-
10^{-120}m_{\rm Planck}^4$, in order for the universe not to collapse 
before 
life has had time to develop[14], and if positive it would have to be 
less than 
about $+10^{-
118} m_{\rm Planck}^4$, in order for galaxies to have had a chance to 
form 
before the matter energy density fell too far below the vacuum energy 
density[15].  
As far as I know, this is at present the only way of understanding 
the small 
value of the vacuum energy.  But of course it makes sense only if the 
big bang 
in which we live is one of an ensemble of many big bangs with a wide 
range of 
values of the cosmological constant.  There are various ways that 
this might be 
realized:

\begin{description}
\item {(a)} Wormholes or other quantum gravitational effects may 
cause the wave 
function of the universe to break up into different incoherent terms,  
corresponding to various possible universes with different values for 
what are 
usually called the constants of nature, perhaps including the 
cosmological 
constant[16].

\item {(b)} Various versions of ``new'' inflation lead to a continual 
production 
of big bangs[17], perhaps with different values of the vacuum energy.  
For 
instance, if there is a scalar field that takes different initial 
values in the 
different big bangs, and if it has a sufficiently flat potential, 
then its 
energy appears like a cosmological constant, which takes different 
values in 
different big bangs[18].  

\item {(c)} As the universe evolves the vacuum energy may drop 
discontinuously 
to lower and lower discrete values.  One way for this to happen is 
for the 
vacuum energy to be a function of a scalar field, with  many local 
minima, so 
that as the universe evolves the vacuum energy keeps dropping 
discontinuously to 
lower and lower local minima[19].  Another possibility[20] with 
similar  
consequences is based on the introduction of an antisymmetric gauge 
potential 
$A_{\mu\nu\lambda}$, which enters in the Lagrangian density in a term 
proportional to  $F^{\mu\nu\lambda\kappa}F_{\mu\nu\lambda\kappa}$, 
where $ 
F_{\mu\nu\lambda\kappa}$ is $\partial_\kappa A_{\mu\nu\lambda}$ with 
antisymmetrized indices.  Instead of a scalar field tunneling from 
one minimum 
of a potential to another, the vacuum energy evolves through the 
formation 
of membranes, across which there is a discontinuity in the value of 
Lorentz-invariant gauge fields $ 
F_{\mu\nu\lambda\kappa}=F\epsilon_{\mu\nu\lambda\kappa}$.  To allow 
an 
anthropic 
explanation of the 
smallness of the vacuum energy, it is essential that the metastable 
values of 
the vacuum energy be very close together.  Several models of this 
sort have been 
proposed recently[21].

\end{description}

Under any of these alternatives, we have not only an  upper bound[15] 
on the 
vacuum energy density, given by the matter energy density at the time 
of 
formation of the earliest galaxies, but also  a plausible 
expectation, which 
Vilenkin calls the principle of mediocrity[22],  that the vacuum 
energy 
density found by typical astronomers will be comparable to the mass 
density at 
the time 
when most galaxies condense, since any larger vacuum energy 
density would reduce 
the number of galaxies formed, and there is no reason why the vacuum 
energy 
density should be much smaller.  The observed vacuum energy density 
is somewhat 
smaller than this, but not very much smaller.  This can be put 
quantitatively[23]:  under the assumption[24] that the {\em a priori} 
probability distribution of the vacuum energy is approximately 
constant within 
the narrow range within which galaxies can form, the probability 
that an 
astronomer in any of the big bangs would find a value of 
$\Omega_\Lambda$ as 
small as $0.7$ ranges from 5\% to 12\%, depending on various 
assumptions about 
the initial fluctuations.  In this calculation the fractional 
fluctuation in the 
cosmic mass density at recombination is assumed to take the value 
observed in 
our big bang, since the vacuum energy would have a negligible 
effect on physical 
processes at and before recombination.  There are also 
interesting calculations 
along these 
lines in which the rms value of density fluctuations 
at recombination is allowed 
to vary independently of the vacuum energy[25].
\vspace*{10pt}

\section{COSMIC MICROWAVE BACKGROUND ANISOTROPIES} 

Perhaps the most remarkable improvement in cosmological knowledge 
over the past 
decade has been in studies of the cosmic microwave background.  Since 
COBE, 
there is  for the first time a cosmological parameter --- the 
radiation 
temperature --- that is known to three significant figures.  More 
recently, 
since the BOOMERANG and MAXIMA experiments reviewed here by Paolo de 
Bernardis, 
our knowledge of small angular scale anisotropies has become good 
enough to set 
useful limits on other cosmological parameters, such as the present 
spatial 
curvature.  

Unfortunately, this has produced a frustrating situation 
for those of us who are 
not specialists in the theory of the cosmic microwave 
background.  We see papers 
in which experimental results for the strengths $C_\ell$ of the 
$\ell$th 
multipole in the temperature correlation function are compared with 
computer 
generated plots of $C_\ell$ versus $\ell$ for various 
values of the cosmological 
parameters, without the non-specialist reader being able to 
understand why the 
theoretical plots of $C_\ell$ versus $\ell$ look the way they do, or 
why they 
depend on cosmological parameters the way they do. I want to take the 
opportunity here to advertise a formalism[26] that I think helps in 
understanding 
the main features of the observed anisotropies, and how they depend 
on various 
cosmological assumptions.

One can show under very general assumptions that the fractional 
variation from the mean of the cosmic microwave background 
temperature observed 
in a direction $\hat{n}$ takes the form
\begin{equation}
\frac{\Delta T(\hat{n})}{T} 
=\int d^3k\,\epsilon_{\bf k}\,e^{id_A{\bf 
k}\cdot\hat{n}}\,\left[F(k)+i\,\hat{n}\cdot\hat{k}\,G(k)\right]\;,
\end{equation}
where $d_A$ is the angular diameter distance of the surface of 
last scattering
\begin{equation}
d_A=\frac{1}{\Omega_C^{1/2}H_0(1+z_L)}\sinh\left[\Omega_C^{1/2}\int_{
\frac{1}{1+
z_L}}^1\frac{dx}{\sqrt{\Omega_\Lambda 
x^4+\Omega_Cx^2+\Omega_Mx}}\right]\;,
\end{equation}
(with $z_L\simeq 1100$ and $\Omega_C\equiv 1-\Omega_M-\Omega_V$); 
${\bf 
k}^2\epsilon_{\bf k}$ is proportional to the Fourier transform of the 
fluctuation in the energy density at early times (with ${\bf k}$ the 
physical 
wave number vector at the nominal moment of last scattering, so that 
$ d_A {\bf 
k} $ in the argument of the exponential is essentially independent of 
how this 
moment is defined); and $F(k)$ and $G(k)$ are a pair of form factors 
that 
incorporate all relevant information about acoustic oscillations up 
to the time 
of last scattering, with $F(k)$ arising from intrinsic temperature 
fluctuations 
and the Sachs--Wolfe effect, and $G(k)$ arising from the Doppler 
effect.  Given 
the form factors, one can find the coefficients $C_\ell$ for $\ell\gg 
1$ by a 
single integration
\begin{equation}
\ell(\ell+1)C_\ell \rightarrow \frac{8\pi^2\ell^3 
}{d_A^3}\int_1^\infty d\beta\, 
{\cal 
P}(\ell\beta/d_A) \left[\frac{\beta F^2(\ell\beta/d_A 
)}{\sqrt{\beta^2-1}} + 
\frac{\sqrt{\beta^2-1}\,G^2(\ell\beta/d_A )}{\beta} \right]\;.
\end{equation}
where ${\cal P}(k)$ is the power spectral function, defined by
\begin{equation} 
\left\langle \epsilon_{\bf k}\,\epsilon_{\bf k'}\right\rangle
=\delta^3({\bf k}+{\bf k'})\,{\cal P}(k)\;.
\end{equation}
(The first term in the square brackets in Eq.~(8) appeared in 
a calculation by Bond and Efstathiou[27]; 
I think 
the second is new.) 

As you can see from the $F^2(k)$ term in Eq.~(8), for $\ell\gg 1$ the 
main 
contribution to $C_\ell$ of the Sachs--Wolfe effect and intrinsic 
temperature 
fluctuations comes from wave numbers close to $d_A/\ell$, but this 
well-known result is not a good approximation for the Doppler effect 
form factor 
$G(k)$.  Since it is the form factors rather than $C_\ell$ that 
really reflect 
what was going on before-recombination, it is important to try to 
measure them 
more directly, as for instance through  interferometric measurements 
of the 
temperature correlation function, of the sort described in a parallel 
session by 
K. Y. Lo {\em et al.} and B. S. Mason {\em et al}.

The Harrison--Zel'dovich spectrum suggested by theories of new 
inflation[28]  
is ${\cal P}(k)=B k^{-3}$, with $B$ a constant.  In this case Eq.~(8) 
gives a 
formula for $C_\ell$ that is valid for $\ell\gg 1$ and $\ell\ll 
d_A/d_H$ 
(where $d_H\ll d_A$ is the horizon distance at the time of last 
scattering):
\begin{equation}
\ell(\ell+1)C_\ell\rightarrow 8\pi^2BF_0^2\left\{1-\frac{ 
\ell^2}{d_A^2}\left[d^2\,\left(\ln\left(\frac{\bar{d}\,\ell}{2d_A}
\right)
-C\right)- d'^2\right]+\dots\right\}\;,
\end{equation} 
where $C$ is the Euler constant $C\equiv -\Gamma'(1)=0.57722$, and 
$d$ and $d'$ 
are a pair of  characteristic lengths of order $d_H$:
\begin{equation}
d^2\equiv \frac{2F_0F_2+G_1^2}{F^2_0}\;,~~~~~~ d'^2\equiv 
\frac{3F_0F_2+\frac{1}{2}G_1^2}{F^2_0} 
\;,
\end{equation} 
expressed in terms of coefficients in a power series expansion of the 
form 
factors:
\begin{equation}
F(k)=F_0+F_2\,k^2+\cdots\;,~~~~~~~~G(k)=G_1k+G_3k^3+\cdots\;.
\end{equation}
(This formula applies even when $\ell$ is not much larger than unity, 
except for $\ell=0$ and $\ell=1$ [29], provided 
we replace $\ell^2$ with $\ell(\ell+1)$ and $\ln \ell$ with 
$\sum_{r=1}^\ell 
1/r+C$.)  The quantity  $\bar{d}$ in the logarithm is another length 
of order 
$d_H$, this 
one given by a much more complicated expression involving the form 
factors at 
all wave numbers, but since $d_H\ll d_A$ the precise value of 
$\ln(\bar{d}/2d_A)$ does not depend sensitively on the precise value 
of 
$\bar{d}$.

One advantage of this formalism is that it  provides a nice 
separation between 
the three different kinds of 
effect that influence the observed temperature fluctuation, that 
arise in three 
different eras: the power spectral function ${\cal P}(k)$ 
characterizes the 
origin of the fluctuations, perhaps in the era of inflation; the form 
factors 
$F(k)$ and $G(k)$ characterize acoustic fluctuations up to the time 
of last 
scattering; and the angular diameter distance $d_A$ depends on the 
propagation 
of light since then.  This allows us to see easily what depends on 
what 
parameters. 
The form factors $F(k)$ and $G(k)$ depend strongly on $\Omega_B h^2$ 
(through 
the effect of baryons on the sound speed) and more weakly on 
$\Omega_M h^2$ 
(through the  effect of radiation on the expansion rate before the 
time of 
last scattering), but since the 
curvature and vacuum 
energy were negligible at and before last scattering, $F(k)$ and 
$G(k)$ are 
essentially independent of the present curvature and of 
$\Omega_\Lambda$.  The 
power spectral function ${\cal P}(k)$ is expected to be 
independent of all these 
parameters.  On the other hand, $d_A$ is affected by whatever 
governed the paths 
of light rays since the time of last scattering,
so it depends strongly on $\Omega_M$, $\Omega_\Lambda$, and the 
spatial 
curvature, but it 
is essentially independent of $\Omega_B$.  In quintessence theories 
$d_A$ would 
be given by a 
formula different from (7), but ${\cal P}(k)$ and the form factors 
would be 
essentially 
unchanged as long as the quintessence energy density was a small part 
of the 
total energy density at and before the time of last scattering.
In particular, Eq.~(8) shows that $\ell(\ell+1) C_\ell$ for $\ell\gg 
1$ depends 
on $\ell$ and $d_A$ only through the ratio $\ell/d_A$, so changes in 
$\Omega_\Lambda$ or the introduction of quintessence would lead to a 
re-scaling 
of all the $\ell$-values of the peaks in the plots of 
$\ell(\ell+1)C_\ell$ 
versus $\ell$, but would have little effect on their height.  

\begin{figure}
\centerline{\epsfbox{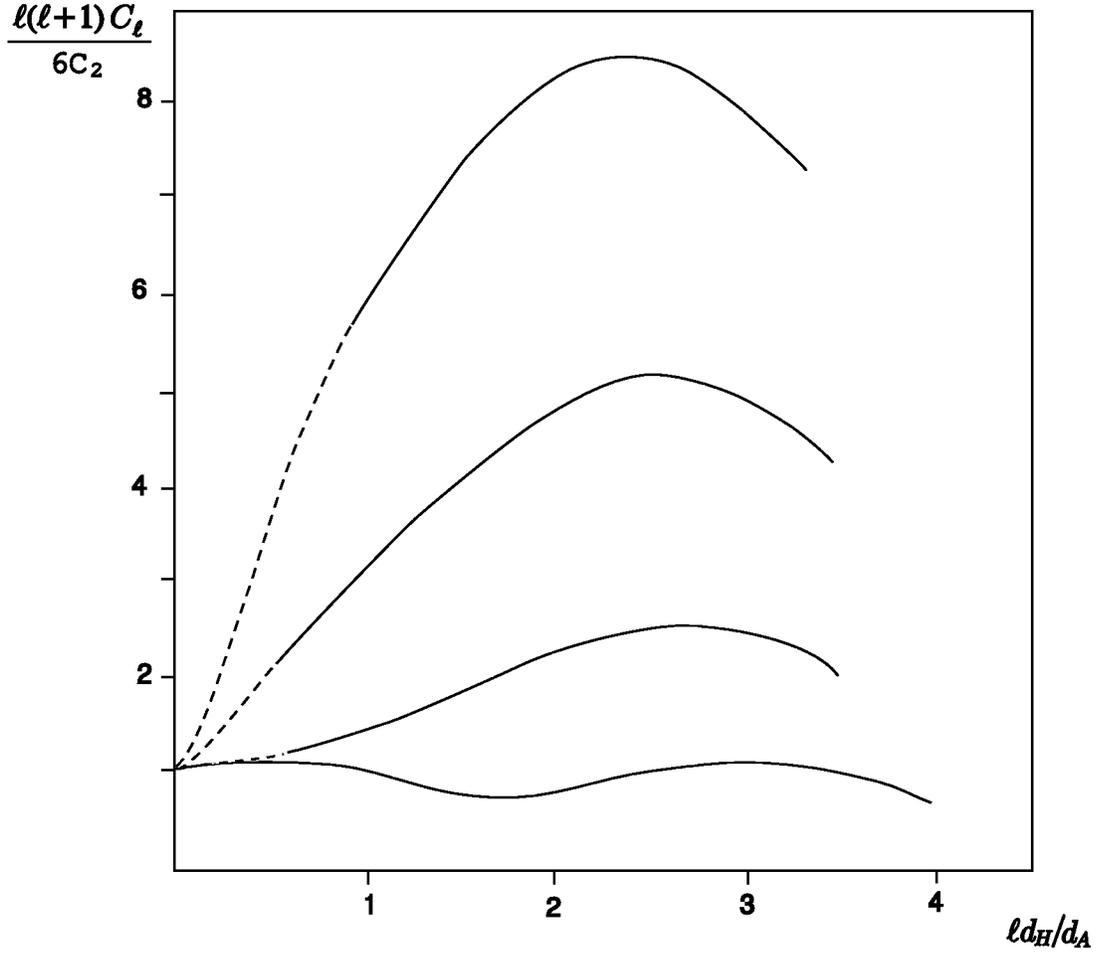}}
\vspace{-300pt}
\caption{Plots of the ratio of the multipole strength
parameter $\ell(\ell+1)C_\ell$
to its value at small $\ell$, versus $\ell d_H/d_A$, where $d_H$ is
the horizon size at the time of last scattering and $d_A$  is the
angular diameter distance of the surface of last scattering.  The
curves are for $\Omega_Bh^2$
ranging (from top to bottom) over the values 0.03, 0.02, 0.01, and 0,
corresponding to $\xi$ taking the values 0.81, 0.54, 0.27, and 0.
The solid curves are calculated using the WKB approximation; dashed 
lines
indicate an extrapolation to the known value at small  $\ell 
d_H/d_A$.}
\end{figure}

Another advantage of this formalism is that, although $C_\ell$ must 
be 
calculated by a numerical integration, it is possible to give 
approximate 
analytic expressions for the form factors in terms of elementary 
functions, at 
least in the approximation that the dark matter dominates the 
gravitational 
field for a significant length of time before last scattering.  
(There have been 
numerous earlier analytic calculations of the temperature 
fluctuations[30], and 
their results may all be put in the form (6), but my point here is 
that this 
form is general, not depending on the particular approximations 
used.)  In this 
approximation the form factors for very small wave numbers are
\begin{eqnarray} 
F(k)&\rightarrow & 1-3k^2t_L^2/2-3[-\xi^{-
1}+\xi^2\ln(1+\xi)]k^4t_L^4/4+\dots\;,\\
G(k)&\rightarrow & 3kt_L-3k^3t_L^3/2(1+\xi)+\dots\;,
\end{eqnarray} 
while for wave numbers large enough to allow the use of the WKB 
approximation
the form factors are
\begin{equation}
F(k)= (1+2\xi/k^2t_L^2)^{-1}\left[-3\xi+2\xi/k^2t_L^2+(1+\xi)^{-
1/4}e^{-
k^2d_\Delta^2}\cos(kd_H)\right]\;,
\end{equation} 
 and 
\begin{equation}
G(k)=\sqrt{3}\,(1+2\xi/k^2t_L^2)^{-1} (1+\xi)^{-3/4} e^{-
k^2d_\Delta^2}\sin(kd_H)\;.
\end{equation}
Here $t_L$ is the time of last scattering; $\xi=27\Omega_Bh^2$ is 
$3/4$ the 
ratio of the 
baryon to photon energy densities at this time;
$d_H$ is the acoustic horizon size at this time; 
and $d_\Delta$ is a damping length, typically less than $d_H$.
Using these results in Eq.~(8) gives the curves for 
$\ell(\ell+1)C_\ell/6C_2$ versus $\ell d_H/d_A$ shown in Figure 1, in 
the 
approximation that damping and the term $2\xi/k^2 t_L^2$ may be 
neglected near 
the peak.  In this approximation the scalar form factor $F(k)$ has a 
peak at 
$k_1 =\pi/ d_H $ for any value of $\Omega_Bh^2$, but the peak in 
$\ell(\ell+1)C_\ell$ 
does {\em not} appear  (as is often said) at $\ell=k_1d_A=\pi 
d_A/d_H$; 
instead, 
$\ell d_H/d_A$ at the peak ranges from 3.0 to 2.6, depending on the 
value of 
$\Omega_Bh^2$.  

We see even from these crude calculations how sensitive is the height 
of the 
first peak in $\ell(\ell+1)C_\ell/6C_2$ to the baryon density 
parameter 
$\Omega_Bh^2$.  (The experimental value[31] for the height of this 
peak is about 6.)  Right now, there is some worry about the fact that 
the value of 
$\Omega_Bh^2$ inferred from the ratio of the heights of the second 
and first 
peaks is larger than that inferred from considerations of 
cosmological 
nucleosynthesis.  Perhaps it would be worth trying to estimate 
$\Omega_Bh^2$ by 
comparing theory and experiment for the ratio of $\ell(\ell+1)C_\ell 
$ at the 
first peak to its value for small $\ell$, discarding the data at the 
second peak 
where the statistics are worse and complicated damping effects make 
the theory 
more complicated.\footnote{At the meeting someone in the audience 
said that 
this 
has been done, but that was in the early days, not I  think with the 
more 
detailed information now available.}  

\begin{center}
{\bf ACKNOWLEDGMENTS}
\end{center}

I am grateful to Willy Fischler, Hugo Martel, Paul Shapiro, and Craig Wheeler
 for their help in preparing this report.  This research was supported in part
by a grant from the Welch Foundation and by National Science Foundation Grants 
PHY 9511632 and PHY 0071512.

\begin{center}
{\bf REFERENCES}
\end{center}

\begin{enumerate}

\item For a review, see J. Polchinski, in {\em Fields, Strings, and 
Duality -- 
TASI 1996}, eds. C. Efthimiou and B. Greene (World Scientific, 
Singapore, 1996): 
293.

\item This was first discussed in the context of string theory by I. 
Antoniadis, 
{\em Phys. Lett.} {\bf B246}, 377 (1990); I. Antoniadis, C. 
Mu\~{n}oz, and M. 
Quir\'{o}s, {\em Nucl. Phys.} {\bf B397} 515 (1993); I. Antoniadis, 
K. Benakli,  
and M. Quir\'{o}s, {\em Phys. Lett.} {\bf B331}, 313 (1994); J. 
Lykken, {\em 
Phys. Rev.} {\bf D54}, 3693 (1996); E. Witten, {\em Nucl. Phys.} {\bf 
B471}, 135 
(1996); and then developed in more general terms by N. Arkani-Hamed, 
S. 
Dimopoulos, and G. Dvali, {\em Phys. Lett.} {\bf B 429}, 263 (1998); 
I. 
Antoniadis, N. Arkani-Hamed, S. Dimopoulos, and G. Dvali, {\em Phys. 
Lett.} {\bf 
B 436}, 257 (1998).  A different approach has been pursued by L. 
Randall and R. 
Sundrum, {\em Phys. Rev. Letters} {\bf 83}, 3370 (1999).

\item N. Arkani-Hamed, S. Dimopoulos, and G. Dvali, {\em Phys. Rev.} 
{\bf 59}, 
086004 (1999); S. Hannestad and G. G. Raffelt, astro-ph/0103201.

\item S. Hannestad, astro-ph/0102290.

\item H. Georgi, H. Quinn, and S. Weinberg, {\em Phys. Rev. Lett.}
{\bf 33}, 451 (1974).

\item S Dimopoulos and H. Georgi, {\em Nucl. Phys.} {\bf B193}, 150 
(1981); J. 
Ellis, S.
Kelley, and D. V. Nanopoulos, {\em Phys. Lett.} {\bf B260},
131 (1991); U. Amaldi, W. de Boer, and H. Furstmann, {\em
Phys. Lett.} {\bf B260}, 447 (1991); C. Giunti, C. W. Kim
and U. W. Lee, {\em Mod. Phys. Lett.} {\bf 16}, 1745 (1991);
P. Langacker and M.-X. Luo, {\it Phys. Rev.} {\bf D44}, 817
(1991).  For other references and more recent analyses of
the data, see P. Langacker and N. Polonsky, {\em Phys. Rev.}
{\bf D47}, 4028 (1993); {\bf D49}, 1454 (1994); L. J. Hall
and U. Sarid, {\em Phys. Rev. Lett.} {\bf 70}, 2673 (1993).

\item S. Dimopoulos, S. Raby, and F. Wilczek, {\it Phys.
Rev.} {\bf D24}, 1681 (1981).

\item K. R. Dienes, E. Dudas, and T. Ghergetta, hep-ph/9806292, 
9807522.

\item For recent detailed reviews, see S. Weinberg, in {\em Sources 
and Detection of Dark Matter and Dark Energy in the Universe --- 
Fourth International Symposium}, D. B. Cline, ed. (Springer, Berlin, 
2001), p. 18; E. Witten, {\em ibid.}, p. 27; and J. 
Garriga and A. Vilenkin, hep-th/0011262.

\item K. Freese, F. C. Adams, J. A. Frieman, and E. Mottola, {\em 
Nucl. Phys.} 
{\bf B287}, 797 (1987); P. J. E. Peebles and B. Ratra, {\em 
Astrophys. J.} 
{\bf 325}, L17 (1988); B. Ratra and P. J. E. Peebles, {\em Phys. 
Rev.}
 {\bf D 37}, 
3406 (1988); C. Wetterich, {\em Nucl. Phys.} {\bf B302}, 668 (1988).

\item C. Armendariz-Picon, V. Mukhanov, and P. J. Steinhardt, 
astro-ph/0004134.

\item N. Arkani-Hamed, S. Dimopoulos, N. Kaloper, and R. Sundrum, 
{\em Phys. 
Lett.} {\bf B 480}, 193 (2000); S. Kachru, M. Schulz, and E. 
Silverstein, {\em 
Phys. Rev.} {\bf D62}, 045021 (2000).

\item J. E. Kim, B. Kyae, and H. M. Lee, hep-th/0011118.

\item J. D. Barrow and F. J. Tipler, {\it The Anthropic
Cosmological Principle} (Clarendon Press, Oxford, 1986).

\item S. Weinberg, {\em Phys. Rev. Lett.} {\bf 59}, 2607
(1987).

\item E. Baum, Phys. 
Lett. {\bf B133}, 185 (1984); S. W. Hawking, in {\em Shelter Island 
II -- 
Proceedings of the 1983 Shelter Island Conference on Quantum Field 
Theory and 
the Fundamental Problems of Physics}, ed. by R. Jackiw {\em et al.} 
(MIT Press, 
Cambridge, 1985); {\em Phys. Lett.} {\bf B134}, 403 (1984); S. 
Coleman, {\em 
Nucl. Phys.} 
{\bf B 307}, 867 (1988).

\item A. Vilenkin, {\em Phys. Rev.} {\bf 
D 27}, 2848 (1983); A. D. Linde, {\em Phys. Lett.} {\bf B175}, 395 
(1986).

\item J. Garriga and A. Vilenkin, astro-ph/9908115.

\item L. Abbott, {\em Phys. Lett.} {\bf B195}, 177 (1987).

\item J. D. Brown and C. Teitelboim, {\em Nucl. Phys.} {\bf 279}, 787 
(1988).

\item R. Buosso and J. Polchinski, JHEP 0006:006 (2000); J. L. Feng, 
J. March-
Russel, S. Sethi, and F. Wilczek, hep-th/0005276.

\item A. Vilenkin: Phys. Rev. Lett. {\bf 74}, 846 
(1995); in {\em Cosmological Constant and the Evolution of the 
Universe}, ed. by 
K. Sato {\em et al.} (Universal Academy Press, Tokyo, 1996).

\item H. Martel, P. Shapiro, and S. Weinberg, {\em Ap. J.}
{\bf 492}, 29 (1998).

\item S. Weinberg, in {\em Critical Dialogs in Cosmology}, ed. by N. 
Turok 
(World Scientific, Singapore, 1997).  Counterexamples in theories of 
type (b) 
are pointed out in reference [18], and the issue is further discussed 
in 
reference [9].

\item G. Efstathiou, {\em Mon. Not. Roy. Astron. Soc.} {\bf 274}, L73 
(1995); M. Tegmark and M. J. Rees, {\em Astrophys. J.} {\bf 499}, 526 
(1998), J. 
Garriga, M. Livio, and A. Vilenkin, {\em Phys. Rev.} {\bf D61}. 
023503 (2000); 
S. Bludman, {\em Nucl. Phys.} {\bf A663-664}, 865 (2000).

\item S. Weinberg, astro-ph/0103279 and 0103281.

\item J. R. Bond and G. Efstathiou, Mon. 
Not. R. Astr. Soc. {\bf 226}, 655 (1987), Eq.~(4.19).

\item S. Hawking,  Phys. Lett. {\bf 115B}, 295 (1982);  A. A. 
Starobinsky, 
 Phys. Lett.  117B, 175 (1982); A. Guth and S.-Y. Pi, Phys. Rev. 
Lett. {\bf 49}, 1110 (1982); J. M. Bardeen, P. J. Steinhardt, and M. 
S. Turner, 
 Phys. Rev. {\bf D28}, 679 (1983); W. Fischler, B. Ratra, and L. 
Susskind, 
 Nucl. Phys. {\bf B259}, 730 (1985).

\item In Eq.~(6) terms are neglected that only affect $C_0$ and $C_1$; 
for these terms, see A. Dimitropoulos and L.P. Grishchuk, gr-qc/0010087.

\item P. J. E. Peebles and J. T. Yu,  Ap. J. {\bf 162}, 815 (1970);
J. R. Bond and G. Efstathiou,  Ap. J. Lett. {\bf 285}, L45 (1984);  
Mon. Not. 
Roy. Astron. Soc. {\bf 226}, 655 (1987); C-P. Ma and E. Bertschinger, 
Ap. J. 
{\bf 455}, 7 (1995); W. Hu and N. Sugiyama,  Ap. J. {\bf 444}, 489 
(1995); {\bf 
471}, 542 (1996).

\item A. H. Jaffe {\it et al.}, astro-ph/0007333.

\end{enumerate}
\end{document}